\documentclass{article}
\usepackage[utf8]{inputenc}
\usepackage[colorinlistoftodos]{todonotes}
\usepackage{xltabular}
\usepackage{lscape}
\usepackage{longtable}
\usepackage{graphicx}
\usepackage{caption}
\usepackage{subcaption}
\usepackage{amsmath}
\usepackage{geometry}
\usepackage[square,numbers]{natbib}
\usepackage{hyperref}
\hypersetup{
    colorlinks=true,
    linkcolor=violet,
    filecolor=magenta,      
    urlcolor=blue,
		citecolor=blue,
    }

\newcolumntype{L}[1]{>{\raggedright\let\newline\\\arraybackslash\hspace{2pt}}m{#1}}
\usepackage{longtable}

\title{Artificial Intelligence For Breast Cancer Detection:  Trends \& Directions}

\author{Shahid Munir Shah$^1$ \and Rizwan Ahmed Khan$^1$ \and Sheeraz Arif$^1$ \and Unaiza Sajid$^1$ }
\date{%
    $^{1}$ Department of Computer Science, Faculty of Information Technology, Salim Habib University, Karachi, Pakistan
    {}
}

\setcitestyle{square}
\begin{document}

\maketitle

\begin{abstract}
Breast cancer is one of leading cause of death among women. Early diagnosis of breast cancer can significantly improve lives of millions of women across the globe. In the last decade, researchers working in the domain of computer vision and Artificial Intelligence (AI) have beefed up their efforts to come up with the automated framework that not only detects but also identifies stage of breast cancer. The reason for this surge in research activities in this direction are mainly due to advent of robust AI algorithms (deep learning), availability of hardware that can train those robust and complex AI algorithms and accessibility of large enough dataset required for training AI algorithms. Different imaging modalities that have been exploited by researchers to automate the task of breast cancer detection are mammograms, ultrasound, magnetic resonance imaging, histopathological images or any combination of them. This article analyzes these imaging modalities and presents their strengths, limitations and enlists resources from where their datasets can be accessed for research purpose. This article then summarizes AI and computer vision based state-of-the-art methods proposed in the last decade, to detect breast cancer using various imaging modalities. Generally, in this article we have focused on to review frameworks that have reported results using mammograms as it is most widely used breast imaging modality that serves as first test that medical practitioners usually prescribe for the detection of breast cancer. Second reason of focusing on mammogram imaging modalities is the availability of its labeled datasets. Datasets availability is one of the most important aspect for the development of AI based frameworks as such algorithms are data hungry and generally quality of dataset affects performance of AI based algorithms. In a nutshell, this research article will act as a primary resource for the research community working in the field of automated breast imaging analysis. 

\end{abstract}

\section{Introduction}
\label{intro}

Cancer is one of the most fatal disease and breast cancer is the most prevalent type of cancer and biggest cause of mortality among women \cite{anastasiadi2017breast}. According to the statistics published by the World Health Organization (WHO), out of 1,350,000 cases of breast cancer, there are 460,000 deaths each year worldwide \cite{cancer2012comprehensive}. Alone in the United States (US), 268,600 cases of breast cancer were reported in 2019, which is the record figure \cite{desantis2019breast, man2020classification}.


Breast cancer occurs because of abnormal growth of cells in breast \cite{mambou2018breast}. The anatomy of the breast is comprised of different blood vessels, connective tissues, milk ducts, lobules, and lymph vessels \cite{mahmood2020brief}. A tumor is formed in milk ducts or lobules when breast tissues grow abnormally and cell division becomes uncontrolled. The developed tumor may either be benign or malignant. Benign tumors are produced because of the minor structural changes in breast and are classified as noncancerous tumors. On the other hand,  malignant tumors are classified as cancerous tumors and are further categorized as invasive (invasive carcinoma) or non invasive (in-situ carcinoma)  \cite{chiao2019detection}. Invasive breast tumors spread into surrounding organs and create complications \cite{cruz2017accurate}, whereas, non-invasive tumors remain confined into their region and do not invade neighboring organs \cite{richie2003breast}. 

Early detection of breast tumor and its correct diagnosis as benign or malignant is critical to avoid its further advancement and complications. This way, a timely and effective treatment may be planned that in turn decreases the mortality rate caused by breast cancer. 

Several imaging modalities are used for breast cancer detection. X-ray Mammography (MG) \cite{moghbel2019review}, Breast Thermography (BT) \cite{moghbel2013review}, Magnetic Resonance Imaging (MRI) \cite{murtaza2019deep}, Positron Emission Tomography (PET), Computed Tomography (CT) \cite{domingues2020using},  3-D Ultrasound (US) \cite{kozegar2019computer} and Histopathology (HP) \cite{saha2018efficient} are some of the popular imaging modalities used to diagnose and detect breast cancer in early stages.   


Among these imaging modalities, breast mammograms are the most commonly used modality  \cite{cheng2003computer, cheng2006approaches, suh2020automated}. Mammograms are low-dose breast X-rays \cite{mohamed2018deep}, which are easy to capture and often used as first test for breast cancer detection \cite{mehmood2021machine}. Mammography is also a popular method for breast cancer screening of large scale population \cite{van2020effect, hong2020effect}. 

Traditionally, for breast cancer detection and diagnosis, radiologists observe breast images manually (through naked eyes) and after the consensus of other medical experts, finalize their decision. Manual inspection of breast images for possible breast cancer detection is a widely used method, however, certain inescapable facts related to manual inspection of images may lead to inaccurate detection and  prolong the diagnosis process, for example:

\begin{enumerate}
   \item 	Unavailability of the experts in remote areas (under-developed countries).
   \item  Unavailability of the experts with sufficient domain knowledge to precisely analyze multi-class images (images with possible multiple ailment characteristics).
   \item  Inspecting large number of medical images on daily basis may be exhaustive and cumbersome practice.
   \item  Subtle nature of the breast tumor and complex structure of breast tissues make manual analysis more difficult.
   \item  Concentration level of medical experts and other fatigues make the diagnosis harder and a time taking process.
\end{enumerate}

All such facts prolong the diagnosis and lead to false positive or false negative outcomes \cite{motlagh2018breast}.  There is always a need for the additional methods to increase efficiency and to decrease false prediction rate. Recently, Artificial Intelligence (AI) technology has made a great progress in the automated analysis of medical images for anomaly detection. The same is true for breast images for possible breast cancer detection \cite{talo2019automated, george2020breast, rodriguez2019stand}. As compared to manual inspection, AI-based automated image analysis avoid tedious and time consuming screening process and efficiently captures valuable and relevant information from the massive image data. 

AI algorithms (discussed in detail in Section \ref{MLalgo}) can be divided in two categories based on how they interpret data / extract information from the images: 

\begin{enumerate}

	\item Algorithms based on handcrafted features, commonly known as conventional AI / conventional Machine Learning (ML) algorithms (term ML refers the study of AI algorithms that learn from experience without being explicitly programmed \cite{samuel1959some}).
	
	\item Algorithms that processes images and extract information from regions that emerge as salient based on mathematical optimization of classification. Such algorithms are commonly characterized as Deep Learning (DL) / Deep Neural Network (DNN) algorithms  (DL is sub field of ML that refers the study of the AI algorithms that learn representations from data with multiple levels of abstraction \cite{lecun2015deep}). The need of handcrafted features for DL algorithms is minimized and mostly non-existent as DL algorithms learn most salient representation of the data without intervention. 

\end{enumerate}


Each category of AI algorithms have shown remarkable progress in breast imaging analysis and breast cancer detection, however, DL algorithms have been found more promising as compared to conventional ML algorithms \cite{burt2018deep, sharma2020conventional} and have proved themselves to be the strong candidate for the ongoing medial imaging research, particularly of breast cancer imaging research.


%

Keeping in view the aforementioned discussion and progress made by AI algorithms in breast cancer detection, in this article, a critical review of breast imaging analysis using AI algorithms is presented. The review critically analyzes AI algorithms applied on different breast imaging modalities and compares their performance.




Following are the contributions of this article:

\begin{enumerate}
    \item This article presents critical analysis of commonly used breast imaging modalities. Limitations and strengths of different imaging modalities are also discussed.
		
		\item Datasets available for different imaging modalities are presented.
		
		\item Detail of popular DL architectures used for breast imaging analysis are presented along with results.
		

\end{enumerate}


The article is structured as follows: Section \ref{modalities} describes imaging modalities used to detect breast cancer. Section \ref{AI} presents the detail of AI algorithms used for breast cancer detection. This sections discusses progress made by conventional ML algorithm (refer Section \ref{MLalgo1}) and DL algorithms (refer Section \ref{MLalgo2}) in the detection of breast cancer. In the last of this section, details on Convolutional Neural Network (CNN) (refer Section \ref{CNN}), a special architecture of DL used to learn data representation in images, is presented. Special emphasis on CNN is given as this architecture is able to achieve state-of-the-art results for breast cancer detection. Finally, Section \ref{conclusion} concludes the article with summary and future research directions.

\section{Medical imaging modalities used for breast cancer detection}
\label{modalities}

\begin{figure}[h]
\centering
\includegraphics[scale=0.55]{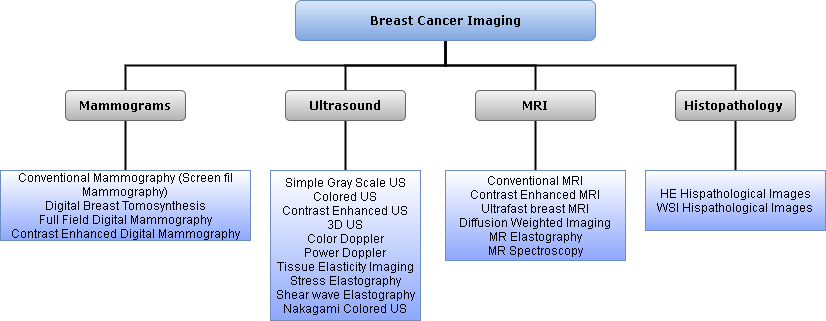}

\caption{Imaging modalities and their sub-types used for breast cancer detection}
\label{Figure-1}
\end{figure}

This section outlines popular imaging modalities used for breast cancer screening and detection. There are various imaging modalities used for this purpose (refer section \ref{intro} for the list of various breast imaging modalities), however, among them four modalities are more commonly used i.e. Mammograms, Ultrasound, Magnetic Resonance Imaging and Histopathology \cite{sree2011breast}. Other than individual use, these modalities are also used in various combinations called multimodalities (Refer Figure \ref{Figure-1}). Detail of each of the above mentioned modalities is provided below.

\subsection{Mammograms}
\label{Mammo}
\begin{figure}[h]
\centering
\includegraphics[scale=0.7]{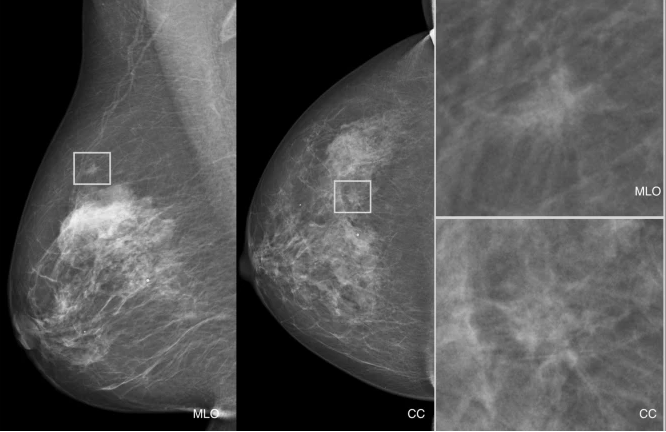}

\caption{Mammographic X-Ray of Human Breast \cite{laang2021identifying}}
\label{Figure-2}
\end{figure}

As stated earlier, mammograms are the most commonly used images for investigating breast tissues for breast cancer detection. These are the low intensity X-ray images of human breast. Figure \ref{Figure-2} shows basic structure of mammograms. 


As indicated in Figure \ref{Figure-2}, in mammograms, cancer tumors and calcium concentration appear brighter as compared to the other masses. Therefore, diagnosis is easy if an experienced radiologist analyze these images or a well trained (trained through mammographic images) ML / DL model is used for the analysis. 

Initially, mammograms were used in their simplest form i.e. Screen Film Mammography (SFM), however, because of technology advancement in images, mammograms were also converted into their advanced forms i.e. Full Field Digital Mammograms (FFDM), Digital Breast
Tomosynthesis (DBT) and Contrast Enhanced Digital Mammograms (CEDM) (refer Figure \ref{Figure-1}). Each category of mammographic images has been widely accepted and used by the research community for breast lesion detection and classification \cite{arevalo2015convolutional, duraisamy2017computer, khan2017automated, hadad2017classification, basile2019microcalcification, kim2016latent, comstock2020comparison}.
One of the important use of mammographic images is to use them for Randomized Mammographic Trials / Randomized Controlled Trials (RMT / RCT). It is a method through which a large scale population is screened for breast cancer detection by analyzing their mammograms. RMT is a globally accepted method and it is the prime reason that mammographic breast images are often considered as the first opinion test for breast cancer detection.

Although, mammography (mammographic images based method) is a popular and commonly used method for screening and detecting breast cancer at early stages \cite{vijayarajeswari2019classification}, in some cases, it is difficult to detect breast cancer at early stages using only mammography and additional screening tests are required along with mammographic trials / randomized mammographic trials \cite{chen2017clarifying}. Especially, in the developing countries where the healthcare infrastructures are deficient and resources are limited, conducting randomized mammographic trials and providing effective treatment are more difficult \cite{da2017breast}. In such cases, breast self-examination (BSE) and clinical breast examination (CBE) are more feasible methods to detect breast cancer at early stages and to reduce mortality. Recent trials have shown that CBE screening successfully reduces diagnosis stages and provide improved breast cancer survival with longer follow-up \cite{yip2018early}. One of the other potential benefit of BSE and CBE is the prevention from a substantial harm related to mammographic screening i.e. over diagnosis magnitude of mammographic X-rays, which is not exactly known.

Other than randomized trials for large population, mammography is also used for detecting breast cancer in individuals. However, in such cases, it is not preferred method because of its limited capability of detecting breast cancer in dense breasted women. Mammographic X-rays often misses to highlight cancerous tissues in young women that have dense breast tissues.  In such cases, Automated Whole Breast Ultrasound (AWBU) / Sonography or other imaging techniques are recommended with mammographic X-rays to acquire a more detailed view of the breast tissues for thorough investigation \cite{cho2017breast}. Table \ref{Table-1} provides more detail about strengths and limitations of using mammograms for breast cancer detection.

\subsection{Ultrasound}
\label{US}

Although mammograms are considered the main modality for early breast cancer detection, these are not safe and certain health risks are associated with them for example, ionizing radiation risks for patients and radiologists and over dosage of radiation risks for patients \cite{fiorica2016breast}.  Furthermore, these modalities lead a large population (65\%-85\%) to unnecessary biopsy operations (refer Section \ref{HP}) because of low specificity \cite{jesneck2007breast, cheng2010automated} (specificity is the test's ability to correctly designate a subject without the disease as negative \cite{maxim2014screening}). Such unnecessary biopsies increase the hospitalization cost for individuals as well as cause mental stress for them. Because of such limitations breast ultrasound imaging are considered much better option for breast cancer detection \cite{zhi2007comparison, han2019reducing}. 

As compared to mammograms, breast ultrasound imaging can increase 17\% overall detection rate and decrease 40\%  overall unnecessary biopsy operations \cite{zhi2007comparison}.  Breast ultrasound images are also known as sonograms in medical terminologies. Sonograms are commonly used for detecting the location of suspicious lesions i.e. Region of Interest (ROI) in breast. For automatically locating lesions, the ultrasound images are used in three broad combinations i.e. simple two dimensional grayscale ultrasound images, color ultrasound images along with shear wave elastography (SWE) added features and Nakagami colored ultrasound images  \cite{murtaza2019deep, youk2017shear}. Elastography in ultrasound images is used to measure tissue stiffness for better differentiation between benign and malignant lesions. Similarly, Nakagami ultrasound images provide additional details of statistical parameters, which are advantageous in extracting ROI. 

\begin{figure}[!htb]
\centering
\includegraphics[scale=0.5]{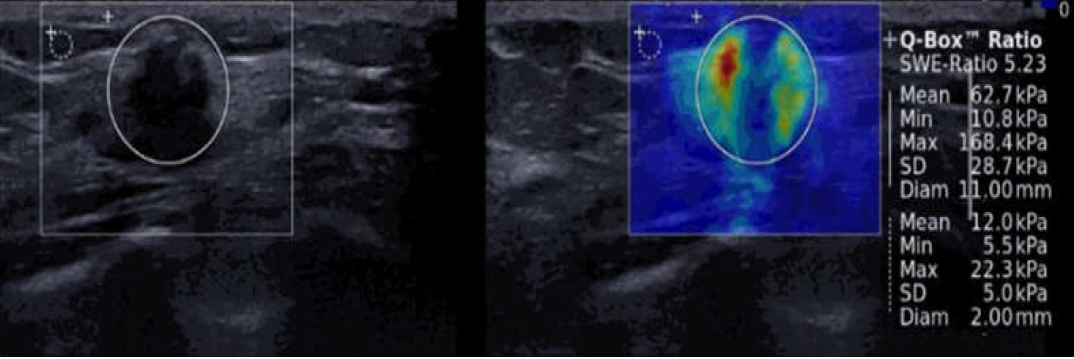}
\caption{Left side: A simple 2D ultrasound image, right side: Shear-wave elastography image \cite{youk2017shear}}
\label{Figure-3}
\end{figure}

\begin{figure}[!htb]
\centering
\includegraphics[scale=0.75]{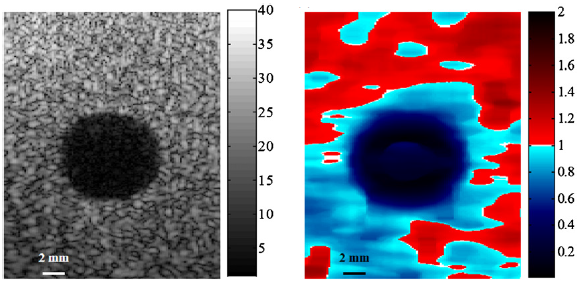}
\caption{Left side: A simple 2D ultrasound image, right side: Ultrasound image with Nakagami distribution \cite{tsui2008classification}}
\label{Figure-4}
\end{figure}

Figure \ref{Figure-3} and Figure \ref{Figure-4} show breast ultrasound images, clearly indicating that a simple breast ultrasound image is just a grayscale image in which irregular mass(es) appears as a big black spot (within the breast mass). However, in color ultrasound images (along with shear wave elastogrphy and Nakagami distribution), irregular masses appear in colors with clear distinguishing boundaries. By analyzing Figure \ref{Figure-3} and Figure \ref{Figure-4}, it can be deduced that colored ultrasound breast images can better detect irregular masses in the breast and can clearly identify the ROI. 

Other than grayscale, SWE and Nakagami ultrasound images, some other breast ultrasound imaging techniques have also been reported in literature i.e. color Doppler, power Doppler and 3D ultrasound images (refer Figure \ref{Figure-1} for the list of some well known breast ultrasound imaging techniques reported in literature for breast cancer detection). Power and color Doppler as well as 3D-imaging are the additional ultrasound techniques, which are used to improve the detection accuracy of ultrasound based breast cancer detection \cite{moustafa2020color, lei2021breast}. Studies have shown that 3D breast ultrasound is robust in identifying up to 30\% more cancers in dense breasted women as compared to mammograms \cite{brem2015screening, thigpen2018role}.

Taking in consideration,  the versatility, safety and high sensitivity, breast ultrasound imaging can be considered the better alternative to mammograms for breast cancer detection \cite{stavros1995solid}. However, breast cancer lesion detection and classification using ultrasound imaging require radiologists' expertise and experience. It is because of the complex nature of ultrasound images and presence of speckle noise \cite{yap2017automated}. Other than complex imaging structure, ultrasound images based screening in asymptomatic women causes unacceptable false positive and false negative outcomes \cite{teh1998role}. Hence, there is a little evidence to support the use of breast ultrasound in breast cancer screening and detection. Table \ref{Table-1} provides more detail about the strengths and limitations of using ultrasound imaging technique for breast cancer detection.

\subsection{Magnetic Resonance Imaging}
\label{MRI}

As discussed earlier that randomized mammographic trials is the common  imaging surveillance method for women to early detect breast cancer, specially the women who are at higher risks of developing breast cancer i.e. having strong family history of breast cancer development. Women with family history are at more higher risk than the others of developing the disease at younger age when the breast density is much higher as compared to older age women. Along with ionizing effects and other health risks, mammographic X-rays provide limitations in detecting breast cancer in dense breasted women i.e. more likely young women \cite{kelly2010breast}. These factors limit the effectiveness of screening by mammography. 

\begin{figure}[h]
\centering
\includegraphics[scale=0.55]{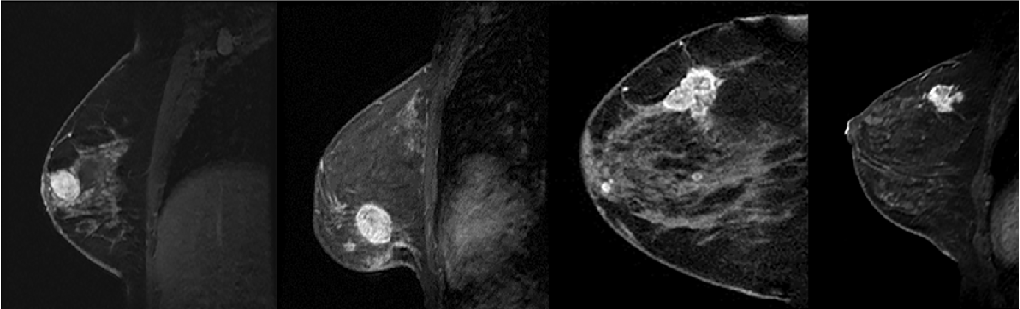}
\caption{Example of breast MRI images \cite{sheth2020artificial}}
\label{Figure-5}
\end{figure}

In contrast breast MRI provides higher sensitivity for breast cancer detection in dense breasts \cite{sardanelli2004sensitivity}. MRI uses magnetic field along with radio waves for capturing clear and detailed images of body soft tissues. It captures multiple breast images of a single subject (taken from different angles) and combines them together as a detailed view, therefore,  breast MRI images provide more detailed view of breast soft tissues than mammograms, ultrasound or computed tomography scanned images \cite{morris2002breast}. Figure \ref{Figure-5} presents a few samples of breast MRI images.  


As indicated in Figure \ref{Figure-5}, MRI images are more detailed images  as compared to the other imaging modalities, therefore,  it may detect lesions, which are not visible on other imaging modalities or could be considered as benign \cite{mann2015breast}. To enhance the image quality of the conventional MRI images (simple grayscale MRI images), usually a contrast agent is injected into the patient body prior to taking MRI images. This technique results in contrast enhanced MRI images, which provide radiologists the opportunity to examine breast tissues in detail with more clear view \cite{rasti2017breast}.    

Since, MRI provide a very detailed view of soft tissues like human breast, it is usually suggested by the doctors or radiologists once the cancer has been diagnosed and further detailed information regarding extent of the disease is needed \cite{mann2008breast}. Sometimes MRI is advised to identify the suspicious breast tissues for biopsy purposes commonly known as MRI guided biopsy. 

Although MRI provide high sensitivity \cite{houssami2018screening}, its use for breast cancer detection is limited because of high associated cost \cite{greenwood2019abbreviated}. However, recently introduced techniques in MRI like Ultrafast breast MRI (UFMRI) and Diffusion-weighted imaging (DWI ) provide significantly higher screening specificity with shorter reading time and reduced overall cost 
\cite{van2018multireader, heller2019mri}. Figure \ref{Figure-1} lists some of the popular MRI methods for breast cancer detection reported in the literature. Table \ref{Table-1} provides details on limitations and strengths of using MRI imaging for breast cancer detection. 



\subsection{Histopathologic Images}
\label{HP}

Histopathology refers to the procedure of taking out a piece of mass from suspicious human body region for testing and detailed analysis by the pathologists \cite{aswathy2017detection}. This process is often termed as biopsy in medical terminologies. For producing histopathology images, the biopsy samples are fixed across the glass slides stained with Haemotoxylin and Eosin (H \& E)  for the examination through an instrument like microscope \cite{tellez2018h}. The purpose of staining with H \& E is to produce colored histopathologic(HP) images for better visualization and detailed analysis of the tissues. Different tissue structures are coloured with different stains for better conceptualization under the microscope. 

\begin{figure}[h]
\centering
\includegraphics[scale=0.55]{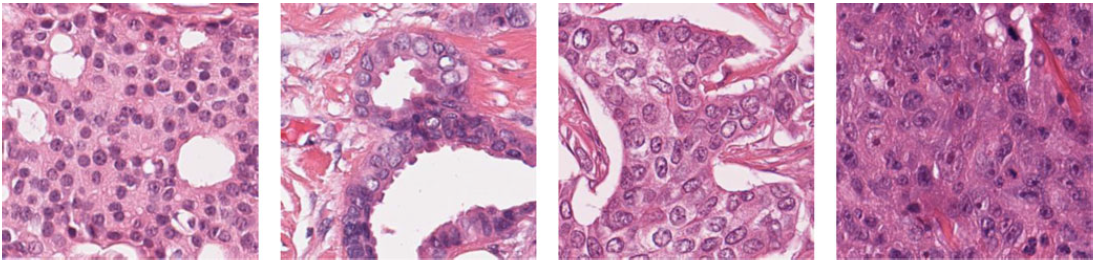}
\caption{Breast histopathologic whole slide images \cite{veta2014breast}}
\label{Figure-6}
\end{figure}

HP images are available in two forms (1) Whole Slide Images (WSI) i.e. digital colored images and (2) image patches extracted from WSI i.e. ROI.  Figure \ref{Figure-6} presents a few samples of HP whole slide images. 


Stained slides are mostly converted in WSIs for the examination by  the expert pathologists. Pathologist extract ROI patches from WSI with different zooming factors to diagnose multiple breast cancer types, which are impossible to diagnose with the help of simple grayscale images. Because of the tissue level examination, multiple researchers have successfully and accurately used HP images for multi-class breast cancer classification \cite{nahid2017histopathological, araujo2017classification, bardou2018classification}. Breast cancer classification using HP images provide various advantages over mammograms and the other imaging modalities like Ultrasound and MRI. For example, HP images are able to classify breast cancer subtypes instead of just binary classes. Furthermore, a large number of ROI can be created using WSI images, which can better train ML / DL models for breast cancer subtype classification. 

Although, HP images provide high authenticity for breast cancer classification especially breast cancer sub type classification, it has some drawbacks. For example, biopsy is an invasive method, requires long time to be converted into digital image form as well as require high expertise to distinguish among breast cancer sub-types. Furthermore, high color variations (because of the staining process), lab
protocols, and scanner brightness in the development of HP images, complicate training a multiclass ML / DL model efficiently, especially when using borderline cases. Refer Table \ref{Table-1} for discussion on strengths and limitations of the HP imaging for breast cancer detection.




\begin{landscape}
\begin{longtable}{L{2.5cm}L{6.5cm}L{6.5cm}L{6.0cm}}   
\caption{Strengths and limitations of different imaging modalities for breast cancer detection}
\label{Table-1}
\\
\hline 
\textbf{Modality} & \textbf{Strengths} & \textbf{Limitations} & \textbf{Available Datasets} \\
\hline 
\endhead
\multicolumn{2}{@{}l}{\ldots \textit{continued on the next page}}
\endfoot
\endlastfoot
Mammograms (MG)
&
\begin{itemize}
\item Mammography is a widely used method for breast cancer screening and detection. 
\item It is an easy and economical first opinion approach for breast cancer diagnosis \cite{jaglan2019breast}. 
\item Digital Mammography (DM) provides an efficient and cost effective solution to capture, store, and process the breast tissue images \cite{posso2017effectiveness}. 
\item Digital Mammographic images serve as a database for training AI systems 
\item DM does not require high expertise or professional knowledge to diagnose and categorize as compared to HP images 
\end{itemize}
& 
\begin{itemize}
\item Mammographic images are produced from the low dose breast x-rays, therefore, provide limited capabilities in  capturing  micro-calcification in human breast because of their extremely small sizes and dispersed shape properties \cite{wilkinson2017microcalcification}.
\item Mammography possess limitations in diagnosing breast cancer in dense breasts by missing cancerous tissues in dense tissues \cite{pisano2005diagnostic}.
\item Mammography is not always accurate in diagnosing breast cancer and therefore some time additional testing may be required for accurate diagnoses \cite{zhao2015limitations}.
\end{itemize}
& 
\begin{itemize}
 \item Mini-MIAS 
 
http://peipa.essex.ac.uk\newline/info/mias.html
 
\item DDSM 

http://marathon.csee.usf.edu\newline/Mammography/Database.html 

\item INBreast 

https://biokeanos.com/source/INBreast

\item BCDR 

https://bcdr.ceta-ciemat.es/information/about

\item CBIS-DDSM

https://wiki.cancerimagingarchive.net\newline/display/Public/CBIS-DDSM

\item MIAS 

https://www.repository.cam.ac.uk\newline/handle/1810/250394 
\end{itemize}
\\
\hline 
Ultrasound (US)
& 
\begin{itemize}
\item Breast US captures breast images in real-time fashion, thus, provide flexibility to view breast lesion from different angles and orientations.


\item US provides reduced chances of false negative diagnosis as it is capable to capture breast images from different orientations.

\item US does not expose patients to any type of harmful radiations, therefore, are considered extremely safe specifically for the pregnant women \cite{rapelyea2018breast}.  
\item US is also considered a safe solution for routine breast cancer screening.
\item US is particularly useful imaging modality for detecting breast cancer in dense breasts where mammography possess limitations \cite{sood2019ultrasound}
\end{itemize}
& 
\begin{itemize}
    \item As compared to mammograms, the quality of the US images is quite poor. 
    \item The quality of the US images  degrades for the denser breasts. 
    \item Breast US can produce misleading diagnosis if the probe of the scanner is not moved or pressed properly \cite{youk2017shear}. 
    \item US waves attenuate in human body muscles, therefore, are not able to clearly display the  tumor contour in breast  \cite{Ultrasound}. 
    \item Extracting ROI for further investigation is difficult from US images.
\end{itemize}
& 
\begin{itemize}
    \item BCDR 
    
    \item BUSI 
    
    https://wiki.cancerimagingarchive\newline.net/pages/viewpage.action?pageId\newline=70226903
    
\end{itemize}

\\
\hline
Magnetic Resonance Imaging  (MRI) 
& 
\begin{itemize}
    \item Like Ultrasound, MRI also do not expose patients to any harmful ionizing radiations. 
    \item MRI images provide a detailed view of internal soft breast tissues and can capture micro classifications. 
    \item MRI can identify suspicious areas that can be further investigated with biopsy, known as MRI guided biopsy. 
    \item  Dynamic Contrast Enhanced MRI (DCE-MRI) imaging provide more clear and detailed view of the soft breast tissues and hence more easily identify the affected breast regions than normal MRI \cite{hodler2019diseases}.
\end{itemize}
& 
\begin{itemize}
    \item MRI is an expensive method as compared to mammograms or ultrasound, therefore, are not commonly used for breast cancer screening.  
    \item Although, MRI provide a very detailed image of the internal breast tissues, still it can miss cancerous tissues that mammograms can detect \cite{reig2020machine}. 
    \item MRI is mostly recommended as a second opinion test after the mammographic test has been conducted. 
    \item MRI is not generally recommended during pregnancy \cite{kalantarova2021pregnancy}. 
    \item In order to enhance MRI images, contrast agents are usually injected, which may create allergies or other complications and hence not recommended for patients especially for kidney patients \cite{garcia2018step}.
\end{itemize}
& 
\begin{itemize}
    \item RIDER Breast MRI
    
    https://wiki.cancerimagingarchive.net\newline/display/Public/RIDER+Breast+MRI
    
    \item Duke-Breast-Cancer-MRI 
    
    https://wiki.cancerimagingarchive.net\newline/pages/viewpage.action\newline?pageId=70226903

\end{itemize}
\\
\hline
Histopathology (HP)
& 
\begin{itemize}
    \item HP images are colored tissue images that have capability to diagnose various type of cancers. 
    \item It is a better prognosis method for early stage breast cancer. 
    \item With the help of HP images, much more in-depth study of breast tissues is possible, therefore, more confident diagnosis of breast cancer is obtained as compared to any other imaging modalities. 
    \item From whole slide HP images, multiple ROI images can be created, which in turns provide more chances to detect cancer tissues thus, reduce FN rate.
\end{itemize}
& 
\begin{itemize}
    \item HP images are taken through breast biopsy, which is an invasive method, therefore, has high associated risks as compared to the other imaging modalities. 
     \item HP images are difficult to analyze and for their accurate analysis, highly experienced, knowledgeable and expert professional pathologists are needed. 
      \item Manual analysis of HP images is a highly time taken process \cite{kumar2020deep}. 
       \item High care is needed during biopsy sample preparation (tissue sample extraction from breast, fixation of tissue samples on microscopic slides and management of color variations originated because of different staining procedures) to decrease the chance of any false diagnosis.
       \item During analysis, HP images must be carefully dealt by the pathologist to accurately diagnose breast cancer. For example, correct slides orientation, pathologist’s attention, and understanding of color variations in stained images \cite{yang2019guided}. 
\end{itemize}
& 

\begin{itemize}
    \item UCI (Wisconsin) 
    
    https://archive.ics.uci.edu\newline/ml/datasets/Breast+Cancer+\newline Wisconsin+(Diagnostic.
    
    \item BICBH 
    
    https://rdm.inesctec.pt/dataset\newline/nis-2017-003
    
    \item BreakHis
    
    https://web.inf.ufpr.br\newline/vri/databases/breast-cancer-\newline histopathological-database-breakhis/

\end{itemize}
\\
\hline
\end{longtable}
\end{landscape}





\section{AI in Medical Image Analysis}
\label{AI}
In recent years, AI as a multidisciplinary approach, has grown nearly into every business to maximize productivity, efficiency, and accuracy \cite{prevedello2019challenges}. Advance computing resources, massive amount of available data and outstanding algorithms' performance have made AI technology more capable and result oriented as compared to before. Apart from healthcare domain it is used in various application areas, i.e. network intrusion detection \cite{NAZIR2021102164}, image synthesis \cite{Crenn2020}, optical character recognition (OCR) \cite{M2m2020}, facial expression recognition \cite{KHAN20131159} etc.

Within the healthcare domain, AI technology is now being used in many important applications like remote patient monitoring, virtual assistance, hospital management and drug discovery etc. \cite{Jalia2020, yu2018artificial}.  Particularly, in medical image analysis and diagnostics, AI is successfully contributing in recognizing complex imaging patterns from the imaging data to provide a better quantitative assessment in an automated and robust manner. Various radiological imaging related tasks i.e.  risk assessment, disease detection, diagnosis or prognosis, and therapy response \cite{giger2018machine} are now being  accomplished more accurately and easily by integrating AI as a tool to assist radiologists and physicians.

\subsection{Why to use AI for medical image analysis}

AI for medical applications use medical imaging (radiology) data to provide better and timely healthcare services \cite{panayides2020ai}. Primary purpose of using AI in medical imaging is to achieve greater efficacy and efficiency in routine clinical practices and to get support in decision making process. 

Through the wide adoption of Electronic Health Records (EHRs) system \cite{shah2020secondary}, radiological imaging data is  continually growing in size and hence, it is now beyond the capacity of the radiologists to analyze such huge amount of imaging data manually. Studies reveal that, in some cases, a  radiologist has to analyze on average one radiology image every 3 to 4 seconds in an 8-hour workday to meet workload requirements \cite{mcdonald2015effects}. Further growth of medical imaging data with disproportionate rate makes this workload requirement even more worse. As the analysis of  radiological imaging data involves high visual perception and robust cognitive abilities, decisions of radiologists related to medical imaging remains uncertain \cite{fitzgerald2001error}. Under unmanageable and constrained workload conditions, uncertainty in decisions further increases. Use of AI for medical imaging analysis decrease the uncertainty and errors in decision making by providing radiologists the opportunity of pre-screened images with identified features. All such human constraints and advancement in both imaging data \& computing resources motivated the healthcare providers to use AI technology in medical imaging field. 

\subsection{AI Algorithms for medical imaging} \label{MLalgo}

\begin{figure}[!htb]
\centering
\includegraphics[scale=0.6]{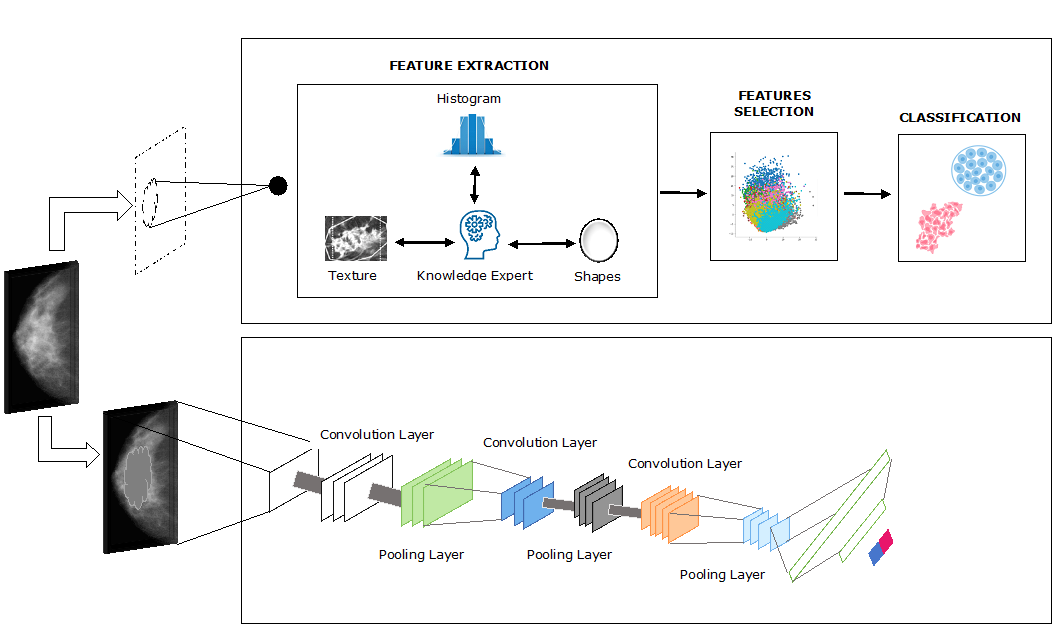}
\caption{ML and DL based approaches for medical imaging data. First row of image denotes standard pipeline used by traditional / conventional AI / Machine Learning (ML) algorithms. Second row of image represents Convolutional Neural Network (CNN) architecture, special case of Deep Learning (DL).}
\label{Figure-7}
\end{figure}

\begin{figure}[!htb]
\centering
\includegraphics[scale=0.6]{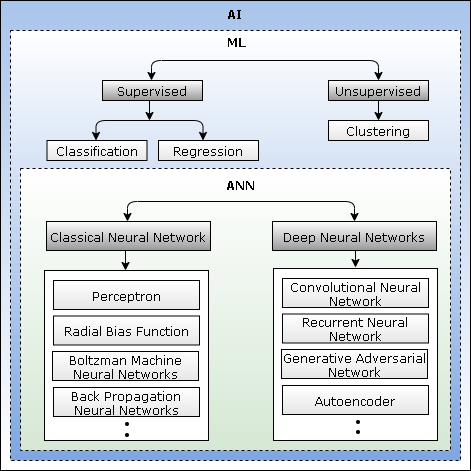}
\caption{Categories of AI algorithms. Acronyms used in the figure: AI= Artificial Intelligence, ML = Machine Learning \& ANN= Artificial Neural Network.}
\label{Figure-100}
\end{figure}

There are two categories of AI algorithms  i.e. the algorithms that use handcrafted features and the algorithms that process raw data. The first category of the AI algorithms belong to the traditional / conventional AI, whereas the the second category of the AI algorithms belong to the recent DL approaches. The same has been shown in Figure \ref{Figure-7}. Refer to Figure \ref{Figure-100}, for graphical representation of types and sub-types of algorithms being employed  for various applications (including healthcare applications) for making or supporting predication. 

\subsubsection{Conventional Machine Learning (ML) algorithms}\label{MLalgo1}

Machine Learning (ML) is a term first coined by Arthur Samuel in 1959 who described it as sub field of AI \cite{samuel1959some}.  It is a technique that recognizes patterns from given inputs, learn those patterns without being explicitly programmed and solve the problems based on inputs \cite{lee2017deep} \cite{erickson2017machine} \cite{muller2016introduction}. In case of medical imaging, manually extracted handcrafted engineered image features (defined in terms of distinguishing image characteristics like area, shapes, region of interest, texture, and histogram of image pixels from the input medical images \cite{tang2019role}) serve as input to ML algorithms. The extracted features may further be processed through features selection algorithms to select most relevant features \cite{NAZIR2021102164} and finally, the task of ML algorithms is to merge the selected input features as a single value such as a tumor signature that might be representing a likelihood of a disease state \cite{clark2013cancer} (refer Figure \ref{Figure-7} for more details).


ML algorithms are categorized into two major types i.e. supervised learning and unsupervised learning algorithms. These two types are further classified into classification, regression (supervised), and clustering (unsupervised) approaches based on the output generated by the algorithms (refer Figure \ref{Figure-100}). 

In supervised learning, the algorithm is trained using labeled data, meaning algorithm learns from prior knowledge / from given probability distribution of classes (in our case prediction of disease). On the other hand, as the name suggests, in the unsupervised learning, the algorithm is not provided with labeled data and it learns pattern from the data and associates them with novel discovered clusters of data points \cite{bazazeh2016comparative, anazirBook}. 
 

Various handcrafted features based ML algorithms have been used for the breast cancer detection and analysis. Yassin et al. \cite{yassin2018machine} in their review have outlined different conventional ML algorithms employed in recent past for the breast cancer diagnosis. These algorithms include but not limited to Decision Tree (DT), Random Forest (RF), Support Vector Machines (SVM), Naive Bayes (NB), K-Nearest Neighbor (KNN), Linear Discriminant Analysis (LDA), and Logistic Regression (LR) \cite{agarap2018breast, sharma2017machine}.

Although, the handcrafted features are considered to be discriminative in nature and ML algorithms based on such features, in some cases, have attained remarkable results in medical imaging domain \cite{azar2013decision, ribeiro2015unsupervised, jian2012computer, kowal2013computer, raghavendra2016application, li2012mammographic, lo2015quantitative, HamRiz}, however, extracting such features is always a challenging task because these features require expert domain knowledge to extract and an exhaustive reprocessing is required to make them suitable as an input to ML algorithms \cite{sigirci2021detection}.

 Furthermore, dependency of such features on expert definitions may miss some contributing factors, which make them imperfect for precise diagnostic process \cite{sharma2020conventional}. Since, the handcrafted features are extracted using feature extracting algorithms (like Gray Level Co-occurrence Matrix (GLCM), Local Binary Pattern (LBP), Graph Run Length Matrix (GRLM), and Histogram of Gradient (HOG) etc.), and these algorithms mostly focus on one aspect of the image (i.e. texture or edge of an image), they may fail with different types of images (for example, a feature extracting algorithm that extracts shape-based features may not be able to extract the features containing texture information). Because of such limitations of extracting inadequate information, systems based on such features are non-adaptable to unknown data (other than the data elements on which the system was trained) and show inability in providing discriminative analysis. Hence, there is always a need of the classifier for making classification decisions to handle the acquired feature space. However, selecting an appropriate classifier is always a complicated task \cite{sharma2020conventional}.

\subsubsection{Deep Learning (DL) based AI algorithms}\label{MLalgo2}

To address the limitations possessed by ML algorithms, recent research is directed towards DL based approaches. DL algorithms process raw data and do not require any manual and explicit extraction of features from the input data to produce output / prediction. These algorithms have the ability to adapt to all kinds of features from input data (image, text, audio, etc.), therefore, provide more robust results in segmentation and classification problems as compared to the conventional ML methods \cite{chan2015pcanet,chen2014deep, chen2017deeplab,he2016deep,ren2015faster}. DL algorithms have also been successfully applied for the evaluation and analysis of medical imaging like CT, MRI, US and HP \cite{cheng2016computer,litjens2016deep,todoroki2017detection}.

\begin{figure}[!htb]
\centering
\includegraphics[width=0.6\textwidth]{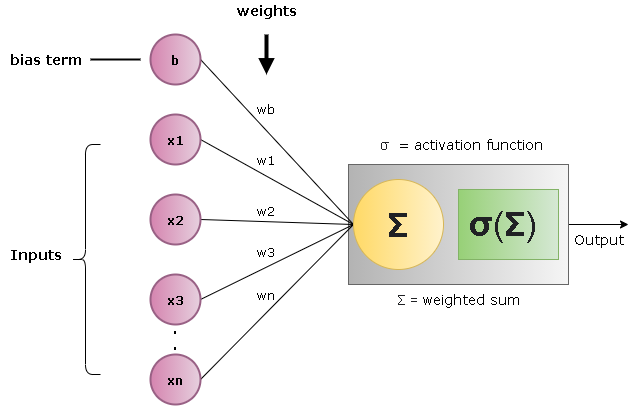}
\caption{Basic Structure of a Neuron.}
\label{Figure-8}
\end{figure}

DL algorithms have been originated from Artificial Neural Network (ANN), which is a subset of ML algorithms (refer Figure \ref{Figure-100} for more detail on taxonomy of algorithms). ANN is the network of artificial neurons that imitates the simple working of brain / biological neurons \cite{king2017guest}. The artificial neuron is the basic building block of ANN. Figure \ref{Figure-8} depicts the basic structure of an artificial neuron, which shows that weighted inputs are provided to a neuron for producing an output or prediction. The neuron sums the received inputs and applies a nonlinear activation function (Sigmoid, Tanh, Gaussian or ReLu etc.) to get the output response (usually between 0 and 1) \cite{Goodfellow-et-al-2016}.

\begin{figure}[!htb]
\centering
\includegraphics[width=0.65\textwidth]{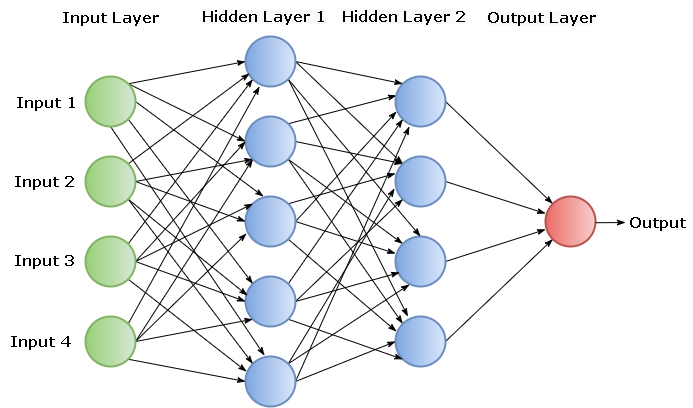}
\caption{A schematic of Illustration of Artificial Neural Network}
\label{Figure-9}
\end{figure}

In ANN, nodes(neurons) are connected to each other in the form of single or multiple hierarchical layers and can send and receive signals as shown in Figure \ref{Figure-9}. The response (rejection or acceptance) of these send and receive signals is dependent upon the nonlinear activation function output. With the input of each neuron or node in a network, a weight is associated that may affect the given input and is useful to transfer the data to output layer. The input layer transmits inputs / data in form of feature vector with a weighted value to hidden layer. The hidden layer, is composed of activation units, carries the features vector from first layer with weighted value and performs calculations as output. The output layer is made up of activation units, each corresponding to label / classes present in the dataset, carrying weighted output of hidden layer and predicts the corresponding class \cite{Goodfellow-et-al-2016}. ANN utilizes the functionality of back-propagation \cite{118638} during training phase to reduce the error function. The error is reduced by updating weight values in each layer.

\begin{figure}[!htb]
\centering
\includegraphics[width=\textwidth]{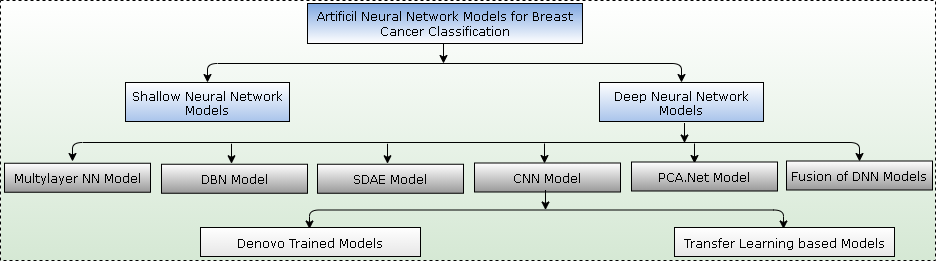}
\caption{Types of Artificial Neural Network used for Breast Cancer Classification.}
\label{Figure-101}
\end{figure}

For breast cancer classification, mainly two types of ANN have been used i.e. Shallow Neural Network (SSN) \cite{bebis1994feed} and Deep Neural Network (DNN) \cite{svozil1997introduction} (refer Figure \ref{Figure-101} for the detail). Among these two neural networks, DNN is the most popular type of ANN that has been extensively used for medical image analysis and diagnosis  \cite{bakator2018deep}. DNN has the ability to automatically extract relevant features from the raw input data without any expert knowledge and any human intervention. Due to aforementioned characteristics, DL approaches provide substantial improvements in diagnostic, analysis and clinical decision-making processes using medical imaging data. 


After getting state-of-the-art results in the domain of image analysis, multiple variants of DNNs have been employed for breast cancer classification in literature. Multi-Layer Network (ML-NN) \cite{ bengio2007learning, arefan2015automatic},  Deep Belief Network (DBN) \cite{ fischer2012introduction, zhang2016deep}, Principal Component Analysis. Net (PCA.Net) \cite{wu2016histopathological}, Stacked Denoising Autoencoder (SDAE) \cite{cheng2016computer}, and various variations of Convolutional Neural Network (CNN) \cite{fukushima1982neocognitron, krizhevsky2012imagenet, WAN201734, Han2017} are some of the variants of DNN that have been used for breast cancer classification using breast imaging data. Every DL algorithm possesses some strengths and limitations when used for medical imaging analysis. Mostly, in literature some variants of CNN, due to robust performance, is used to analyze breast images for cancer detection purpose.

Table \ref{Table-2} summarizes strengths and limitations of DNN models when employed for medical imaging analysis.

\pagebreak


\begin{longtable}{|L{1.1cm}|L{3.5cm}|L{5.5cm}|L{5.5cm}|} 
\caption{Strengths and limitations of DL algorithms used for medical imaging analysis}
\label{Table-2}
\\ \hline
 
\textbf{Model} &\textbf{Description} &\textbf{Advantages} &\textbf{Limitations} \\ \hline
SNN \cite{bebis1994feed} & Single hidden layer feed forward network & 
 \begin{itemize}
\item Due to small network size fewer computing resources and less training time and memory required.
\item Optimization of hyper-parameters for training is not very challenging.
\item Ability to produce reasonable performance in case of small dataset.
\item Generalization performance can be increased by increasing hidden layers.
\item Good performance on low dimensional data 
 \end{itemize}

& 
\begin{itemize}
\item Poor performance on high dimensional data.
\item Low performance to solve multi-class problems.
\item Not easy to generalize the predictive result.
\item Offer computational complexity due to large network and many hyper-parameters.
\item Optimization of hyper-parameters for training is very challenging.
\item Requirement of large datasets to train and attain good performance.
\end{itemize}
\\ 
\hline

SDAE \cite{cheng2016computer} & This model extract discriminant representative hidden patterns from data using intrinsic data reconstruction method & 
\begin{itemize}
\item Possess noise reduction ability.
\item Discriminative hidden pattern can be extracted using data reconstruction mechanism.
\item Easy regularization and optimization of training parameters.
\item Auto noise elimination ability helps to extract relevant features.
\end{itemize}
 &
\begin{itemize}
\item Poor performance on low dimensional data.
\item Possessing poor correlation among the dimensions 
\item Auto noise reduction property sometimes not good for low dimensional data 
\end{itemize}
\\ 
\hline

DBN \cite{ fischer2012introduction, zhang2016deep}   & Generative model comprises of several layers that follow greedy layer wise feature learning and training. All hidden layers are trained one layer at a time. It can be work as semi-supervised learning & 
\begin{itemize}
\item Layer-by-layer training practice helpful to improve feature generalization.
\item Easy to optimize hyper-parameter of each layer.
\item Layers of the network can be trained in un-supervised manner which is very demanding in BrC classification.
\item Show better training performance in case of small annotated images.
\end{itemize}
& 
\begin{itemize}
\item DBN cannot track the loss.
\end{itemize}
\\
\hline
CNN (De novo) \cite{WAN201734} & Different layers (input, convolutional, pooling, fully-connected layers) are hierarchal arrange in customizable manner. Model is trained from scratch in supervised manner.
& 
\begin{itemize}
\item Customized model can be developed according to type and number of images. 
\item Preferred when source images are not enough for training.
\end{itemize}
& 

\begin{itemize}
\item In case of datasets of different domains, it is difficult to optimize the model training.
\item Hard to solve multi-class problems if available number images are small 
\end{itemize}
\\
\hline
CNN (Pre-trained) \cite{Han2017} & Model is trained using TL (transfer learning) using pre-trained available networks. Network’s layers are same as CNN (De novo) such as input, convolutional, pooling and FC-layers but in different hierarchal arrangement. 
& 
\begin{itemize}
\item Model can be trained quicker than de novo in case of less available resources.
\item Show reasonable performance in case of small target data.
\end{itemize}
& 
\begin{itemize}
\item Produces unreliable results in case of very small datasets
\item Hard to handle and optimize in case of newly appended layers 
\end{itemize}
\\
\hline

\end{longtable}


Among the aforementioned DL architectures, CNN is the most powerful, effective and extensively used for medical imaging analysis particularly breast imaging analysis \cite{yari2020deep}. In the subsequent subsections, CNN architecture is discussed in detail in connection with the breast imaging analysis and breast cancer detection.

\subsubsection{Convolutional Neural Network (CNN) for breast imaging analysis}
\label{CNN}

As discussed,  CNNs are sub-type of deep, feed forward neural networks that have shown robust results for applications involving visual input i.e image \cite {KHAN201961}. Unlike a typical ANN, CNN architecture can take entire data (i.e. image) as an input. In order to train CNN architectures, usually two techniques are used i.e., transfer learning (TL) and de novo (refer Figure \ref{Figure-101} for the detail). In de novo technique, CNN architecture is trained from the scratch and different trained models are combined to obtain the optimal model, while, in transfer learning method, it adopts the pre-trained models for analysis \cite{KHAN201961}. Transfer learning is a convenient method, however, it mostly deals with limited or small datasets \cite{liu1999ensemble}.

\begin{figure}[!htb]
\centering
\includegraphics[scale=0.5]{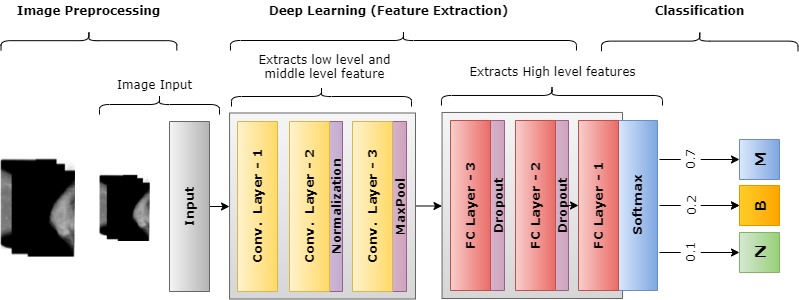}
\caption{The architecture of deep CNN applied to breast cancer detection and classification \cite{murtaza2019deep}. Acronyms used in the figure: Conv. Layer= Convolutional layer, MaxPool= Maximum value pooling and FC Layer= Fully connected layer. Dropout layer is used to reduce overfitting the data. Fully-connected layers are used after multiple convolutional and pooling layers. All neurons of these layers are connected to all neuron of next layer. It helps to aggregate low level features / information to extract large patterns from input data. Softmax function is used in the last layer to calculate probability of occurrence of an event that is present in the training data i.e. malignant or benign tumor or normal breast image (without any tumor).}
\label{Figure-10}
\end{figure}

CNN architecture comprises of hierarchical layers with trainable filters (convolution operation) and pooling operations (reduce the size of the image representation) that capture discriminative features from raw input data to enhance classification performance. Lastly, classification layer computes the probability / score of learned classes from the extracted features \cite{lecun2015deep}. The illustration of deep CNN (DCNN i.e. CNN with many convolutional and pooling layers) applied to breast images classification (for breast cancer detection) is provided in Figure \ref{Figure-10}.


As shown in Figure \ref{Figure-10}, for analysis of breast imaging (or any other medical imaging), the input layer in CNN architecture accepts an input image. Convolutional layers are very important layers that contain convolutional filters known as kernels. These layers use convolution operations to capture the high-level features such as edges, colors, shapes and blobs. The number of layers and convolutional filters are hierarchically arranged in such a way that level of extraction from low-level features to high-level features increases as the depth of layers increases.

\subsubsection{Popular CNN architectures for breast imaging analysis}

  

Studies have revealed that CNN-based models have enhanced the diagnosis accuracy and reduced the false detection rate when used for cancer detection using breast imaging data \cite{wang2019breast}. Using different combinations of parameters and  hyper-parameters (context of depth model), various Transfer Learning (TL) based CNN architectures have been employed for breast imaging analysis such as LeNet \cite{chollet2017xception}, AlexNet \cite{krizhevsky2012imagenet}, GoogLeNet \cite{szegedy2015going}, VGG \cite{szegedy2017inception}, CiFarNet \cite{shin2016deep}, Inception \cite{he2016deep}, Inception v4 \cite{simonyan2014very} and ResNet \cite{szegedy2016rethinking} etc. All such CNN architectures have reported robust results when applied on different breast imaging datasets.

Table \ref{Table-3} presents summary of selected recent articles, published in last 4-5 years, that have used deep learning methods (particularly CNN based frameworks) to detect breast cancer. One commonality of all these articles is that almost all (except few) have presented results using mammograms dataset to prove robustness of proposed framework. 

Further, Table \ref{Table-3} provides detailed information on CNN based architectures / frameworks in terms of training data, network architecture, and performance assessment using different evaluation matrices. Most of the existing schemes are proposed with different arrangements of input size, network depth, and the number of filters. The training and testing datasets utilized by such models may be public or private. 

\begin{landscape}
\begin{longtable}{L{2cm}L{0.8cm}L{2.5cm}L{2.5cm}L{2.5cm}L{2.5cm}L{1.5cm}L{2.5cm}}   
\caption{Performance analysis of different variants of deep learning based models for breast cancer detection}
\label{Table-3}
\\ \hline
\textbf{Research   Team} & \textbf{Year} & \textbf{Task Performed} & \textbf{Data Source} & \textbf{\# of Images} & \textbf{ANN Type and Architecture} & \textbf{Size of Input} & \textbf{Evaluation Results} \\ \hline 

\endhead
\multicolumn{2}{@{}l}{\ldots \textit{continued on the next page}}

\endfoot

\endlastfoot


Fonseca et al. \cite{fonseca2015automatic} & 2015 & Classification: Leision Density & Medical centres in Lima, Peru & 1157 mammograms & CNN: 3Conv Layer + SVM as classifier & 200 x 200 & Area under the   curve : 0.73 \\ 
\hline
Carneiro et al. \cite{carneiro2015unregistered} & 2015 & Classification: mass tissue & DDSM and Inbreast & 680 images (DDSM) and 410 (Inbreast) & CNN: 4 Conv Layer+ 2 FC + Softmax Classifier & 264 x 264 & Area under the   curve : 0.91 \\ 
\hline

Z. Jiao   et al. \cite{jiao2016deep} & 2016 & Classification: benign and malignant & DDSM & 600 [300 each of benign andmalinant] & DeepCNN: 1 input + 5 Conv Layer + 3 Fully Connected & 227 x 227 & Accuracy : 0.967  \\

\hline

Huynh et al. \cite{huynh2016digital} & 2016 & Classification: Benign and Malignant & FFDM & 219 & CNN: 5 Conv Layers + 3 FC + SVM Classifier & 256 x 256 & Area under the   curve : 0.81 \\ 
\hline
Suzuki et al. \cite{suzuki2016mass} & 2016 & Classification: mass and normal region & DDSM & 198 & DCNN: 3 Conv Layer+2 FC+ Softmax Classifier & 224 x 224 & SN: 89.0 \\ 
\hline
Jiao et al.  \cite{jiao2016deep} & 2016 & Mass classification: benign and malignant & DDSM & 600 & CNN: 5 Conv Layers + 2 FC + SVM Classifier & 227 x 227 & Accuracy: 0.967 \\ 
\hline
samala et al. \cite{samala2016mass} & 2016 & Mass detection & DBT volumes from University of South Florida & 2282 digitised films and mammograms & CNN: 4 Conv Layer+ 3 FC & 128 x 128 & Area under the curve : 0.80 \\ \hline

J. Arevalo et al. \cite{arevalo2016representation} & 2016 & Classification: benign and malignant & BCDR- F03  & 736 (426 benign, 310 malignant lesions)
 & CNN:  2 Conv Layer+1 FC+ Softmax Classifier & 150 x 150 & AUC: 0.82 \\ 
\hline

K. K. Smala   et al. \cite{samala2017multi} & 2017 & Classification: benign and malignant & SFM, DM & 2242 (1057 malignant, 1397 benign) & DeepCNN: 5 Conv Layer + 5 Fully Connected & 256 x 256 & Area under the curve : 0.82  \\

\hline

I. Kumar   et al. \cite{kumar2017classification} & 2017 & Classification: Breast Density & MIAS, CBIS-INBreast & 480 MLO view digitized screen film  & CNN & 128 x 128 & Accuracy = 0.57, Area under the curve : 0.77 \\
\hline

Kooi et al. \cite{kooi2017large} & 2017 & Detection of leisions & FFDM & 45000 images (44090 for testing + 18182 for Testing & DCNN: 5 Conv+2 FC + Softmax Classifier & 250 x 250 & Acc: 0.85 \\ 
\hline
Li et al. \cite{li2017deep} & 2017 & Classification into benign and malignant & FFDM & 456 & CNN: 3Conv Layer+2 FC + SVM as classifier & 224 x 224 & Auc: 86.00 \\
\hline
Ahn et al. \cite{ahn2017novel} & 2017 & Classification: non-dense, dense & FFDM & 397 & CNN: Transfer Learning based CNN model & 256 x 256 & Corelation Coeff.:0.96 \\ 
\hline
Dhungel et al. \cite{dhungel2017deep} & 2017 & Classification: Benign and Malignant & Procas & 410 & Cascaded deep learning model: Lenet+RNN & 40 x 40 & Acc: 85.0, SP: 70.0, SN: 98.0 \\ \hline

Mugahed et al. \cite{al2018fully} & 2018 & Classification: Benign, normal and Malignant & Inbreast & 410 & DCNN: Alexnet & 227 x 227 & Acc:89.91, AUC: 94.78, SN:   95.64, F-score: 96.84 \\ \hline
Wu et al. \cite{wu2018breast} & 2018 & Classification: non-dense, dense & Private & 200000 & Transfer Learning based DNN Model & 224 x 224 & Auc: 93.40 \\ \hline
Xu et al. \cite{xu2018classifying} & 2018 & Classification: Low dense and high dense tissue & Inbreast & 410 & Resnet:36 weighted layers, 3 stages with 7 residual learning block & 224x224  & Acc: 96.80 \\ \hline
Zhu et al. \cite{zhu2018adversarial} & 2018 & Classification: Malignant, Benign & DDSM-BRCP & 316 & Fully Convolutional Netwok + CRF:4 FCN (each with 3 layers with multi-scale kernels) & 40x40 & Dice Score: 0.97 \\ \hline

Y. ShaoDe et al. \cite{yu2019transferring} & 2018 & Classification: breast lesion
diagnosis & BCDR-F03  & 736 ((230 benign and 176 malignant) & SCNN: 1 Conv+ 1 Pool+ 1 FC & 128 x 128 & Acc: 73.0, AUC: 0.82 \\ \hline

D. Ribli et al. \cite{ribli2018detecting} & 2018 & Classification: Malignant, Benign & INbreast & 410 ((masses, calcifications, asymmetries, and distortions) & Faster R-CNN: VGG16 & 2100 x 1700 &  AUC: 0.95 \\ \hline

M. A. Al-Masni et al. \cite{al2018automatic} & 2018 & Classification and Detection Simultaneously: Malignant, Benign & DDSM & 600 (benign and malignant) & CNN: Conv24+ Pool1+ FC2 & 7 x 7 &  Acc: 97.0 \\ \hline 

H. Chougrad et al. \cite{chougrad2018deep} & 2018 &  mass lesion classification & DDSM & 5316  (benign and malignant) & InceptionV3: Conv5+ Pool3 & 224x224 &  Acc: 97.35, AUC: 0.98 \\ \hline

F. F. Ting et al. \cite{ting2019convolutional} & 2018 & Classification: Benign and Malignant and Normal & MIAS & 221 (21 benign, 17 malignant, 183 normal) & CNN: 5 Conv+2 FC & 128 x 128 & Acc: 90.5, SP: 90.71, SN: 89.47, AUC: 0.901 \\ \hline

K. Mendal et   al. \cite{mendel2019transfer} & 2019 & Classification: Benign and Malignant & FFDM & 78 & CNN: VGG19 & 224 x 224 & AUC:0.81 \\ 
\hline

Basile et   al. \cite{basile2019microcalcification} & 2019 & Classification: Benign and Malignant & BCDR & 364 & Deep Lerning method & 200 x 200 & SN:92.89 \\ 
\hline
Ragab et al. \cite{ragab2019breast} & 2019 & Classification: Benign, normal and Malignant & DDSM & 5257 images & DCNN: (Alexnet[5 Conv+2 FC] + SVM Classifier) & 227 x 227 & Acc: 87.20, AUC: 94.0 \\ 
\hline

Ionescu et   al. \cite{ionescu2019prediction} & 2019 & Classification: Cancerous, non- cancerous & PROCAS & 73128 & DCNN:4-conv2D layer with 1-fc layer & 640 x 512 & AUC: 61.00 \\ 
\hline

Singh et al. \cite{singh2020breast} & 2020 & Shape Classification: Irregular, lobular, oval, round & Inbreast, DDSM & 410 (inbreast), 1168 (DDSM) & GAN+CNN: 3 Conv Layer+2 FC+ Softmax Classifier & 64 x 64 & Acc: 83.0 \\ 
\hline

Costa et al. \cite{junior2020novel} & 2020 & Classification: Detection of architectural distortion & Private FFDM & 280 & VGG-16:13-Con2D,4 pooling, 3 Dense & 224 x 224 & AuC: 0.89 \\
\hline
Abhijeet et al. \cite{beeravolu2021preprocessing} & 2021 & Classification: Normal, benign, malignant & MIAS & 322 & DCNN+RPN:4-con2d (Dense+Relu+batchNorm+Dropout)+Softmax  & 256 x 256 & Dice Score: 0.97 \\

\hline
altaf et al.\cite{altaf2021hybrid} & 2021 & Classification: Normal, abnormal benign, abnormal malignancy & DDMS, Inbreast, BCDR & 900(DDMS), 300(Inbreast), 450 (BCDR) & Transfer Learning Based PCNN (Pulse-Coupled Neural Network)+DCNN & 224 x 224 & Acc: 98.72 (DDMS), 97.5 (InBreast), 96.30  \
\\ \hline

A. Khamparia et al \cite{khamparia2021diagnosis} & 2021 & Classification: Normal, abnormal benign, abnormal malignancy & DDMS patch & 10713  & MVGG16 + 
ImageNet & 224 x 224 & Acc: 94.3, AUC: 0.933  \
\\ \hline

\end{longtable}
\end{landscape}

 If we analyze the Table-3, we can conclude that different researchers utilized the deep convolutional neural network (DCNN) to perform different tasks for the breast cancer analysis as DCNN architectures achieve state-of-the-art results.  Some research studies focused on the classification of structure and geometry of breast tissues such as dense, non-dense, irregular, lobular, oval, and round shape. Some of the researchers classified the abnormal tissue structure as malignant and benign.  We can also notice from the table that the performance of the DNN mostly depends on the deep architecture, availability of large datasets, high computing resources, and large training input data dimensions. Though, DNN achieved remarkable results in the domain of medical image analysis, however, availability of annotated large medical image datasets and computing resources are still the main requirements.

\section{Conclusion \& Future Directions}
\label{conclusion}

Breast cancer is a critical public health problem and one of the major cause of mortality in women. Early diagnosis and detection, proper control mechanism and cure is necessary to reduce the mortality caused by this deadly disease. Several popular imaging modalities like Mammograms, Ultrasound, Magnetic Resonance Imaging and Histopatholic images among many others are used for breast cancer diagnosis. Traditionally, pathologists’ / radiologists’ observe breast images manually and with the consensus of the other medical experts, finalize their decisions. However, observing large number of breast images manually, for possible breast cancer diagnosis is a cumbersome and time taken process, which often leads to false positive or false negative outcomes. Hence, there is always a need of an automated system to speed up the image analysis process and to help radiologists’ in early diagnosing breast cancer. Such automated systems provide radiologists the opportunity of second opinion and hence, they can make more strong, reliable and accurate decisions regarding breast cancer diagnosis.

Keeping in view the importance of the automated systems for the diagnosis of breast cancer using breast imaging, here in this research study, we have provided a complete road map for the readers to fully understand the working mechanism of AI based automated breast cancer detection systems. We begin our research form describing the basic and the most important imaging modalities, which are being extensively used for breast cancer diagnosis. Along with the comprehensive detail of each imaging modality, its strengths and limitations have also been provided to give readers a broader idea about these imaging modalities. Furthermore, some of the popular available data sets of each modality have also been outlined along with their links so that in case of further research readers can easily access the relevant databases through the given links.

In order to design AI based automated systems for breast imaging analysis, the basic understating of AI algorithms is essential. In this research work, we have provided the basic understanding of AI algorithms. Since, AI algorithms have been broadly divided into handcrafted features based algorithms (conventional AI / ML algorithms) and those, which learn features representations from the raw input data automatically (Deep Learning (DL) algorithms). Both the categories of the AI algorithms have been described in detail to develop the reader’s understanding about AI algorithms. 
Deep Learning algorithms are more popular among the research community to analyze medical imaging data, therefore, an extensive insight of these algorithms has been provided along with their strengths and limitations in connection with breast imaging analysis.

Finally, we diverted our attention towards Convolutions Neural Network (CNN), which is one of the most popular and frequently used DL architecture for medical imaging analysis (particularly breast imaging analysis). Along with the theatrical detail of CNN, this research provides the comprehensive detail of the most recently employed CNN based architectures for breast imaging analysis along with the detail of the data sets used and results obtained by the CNN architectures using these data sets.

Although, AI is playing a significant role in the development of reliable automated systems for medical image analysis and disease predictions, there are still many issues to be addressed before AI can eventually influence clinical practices. One of the main issues is the limited availability of comprehensive and fully labeled datasets and the need for solid ethical regulations. In addition, AI algorithms (particularly DL algorithms) are “black box” in nature i.e. there is no proper justification or explanation of the decisions / predictions made by these algorithms. In medical diagnosis (like breast cancer diagnosis), black box decisions (like the recommendations / decisions made by the AI based automated systems) are usually not preferred as the radiologists’ / physicians’ are mainly interested to know and understand how the decision was made and on what factors it was taken \cite{moxey2010computerized}. Based on the less explainable nature and several other factors like losing control over autonomous decision-making, the automated AI based systems for disease prediction / diagnosis (decision support systems) are often regarded as threat or loss of control by the physicians \cite{liberati2017hinders}.

Explainability of the AI algorithms was well ranked by the physicians many years ago \cite{ teach1981analysis} by considering it one of the most important feature of the AI based automated decision support systems. Today, many researchers like Villani report on artificial intelligence in France \cite{ villani2018donner} and many others recommend “open the black box of artificial intelligence”, and to focus on the use of interpretable models for making high stakes decisions \cite{rudin2019stop}.

Hence, in order to gain the physicians’ confidence on the decisions made by AI algorithms and to justify the reliability of the decisions, a proper explanation of the decisions is needed specifically when these algorithms are used for predicting a disease. This also opens up the research area of ``Ethical AI''. It is therefore the ultimate responsibility of the research community to make the AI algorithms fully explainable and interpretable so that these systems could be regarded as the strong candidates of affecting decision making for possible disease predictions. It will help in widely embedding AI technology in clinical care applications.

\bibliographystyle{unsrtnat}
\bibliography{references}
\end{document}